 \documentclass[sn-basic]{sn-jnl}
\usepackage{bbm}
\usepackage{lscape}
\usepackage{moreverb}
\usepackage{graphicx}%
\usepackage{multirow}%
\usepackage{subcaption}
\usepackage{amsmath,amssymb,amsfonts}%
\usepackage{amsthm}%
\usepackage{mathrsfs}%
\usepackage[title]{appendix}%
\usepackage{xcolor}%
\usepackage{color, soul}
\usepackage{textcomp}%
\usepackage{manyfoot}%
\usepackage{booktabs}%

\usepackage{natbib}
\usepackage{listings}%
\usepackage{styles}
\usepackage{anyfontsize}
\begin{document}
\title[Title]{Using Subject-Level Variability to Predict Time-Varying Outcomes: Investigating the Association between Hormone Variability and BMD Trajectories over the Menopausal Transition}

\author*[1]{\fnm{Irena} \sur{Chen}}\email{chen@demogr.mpg.de}
\author[2, 3]{\fnm{Zhenke} \sur{Wu}}
\author[4]{\fnm{Sioban D.} \sur{Harlow}}
\author[4]{\fnm{Carrie A.} \sur{Karvonen-Gutierrez}}
\author[4]{\fnm{Michelle M.} \sur{Hood}}
\author[2,5]{\fnm{Michael R.} \sur{Elliott}}
\affil*[1]{\orgdiv{Department of Digital and Computational Demography}, \orgname{Max Planck Institute for Demographic Research}, \orgaddress{\city{Rostock}, \country{Germany}}}
\affil[2]{\orgdiv{Department of Biostatistics}, \orgname{University of Michigan},\orgaddress{\city{Ann Arbor},  \state{Michigan}, \country{USA}}}
\affil[3]{\orgdiv{Michigan Institute for Data and AI in Society (MIDAS)}, \orgname{University of Michigan}, \orgaddress{\city{Ann Arbor}, \state{Michigan}, \country{USA}}}
\affil[4]{\orgdiv{Department of Epidemiology}, \orgname{University of Michigan}, \orgaddress{\city{Ann Arbor}, \state{Michigan}, \country{USA}}}
\affil[5]{\orgdiv{Department of Survey Methodology}, \orgname{University of Michigan}, \orgaddress{\city{Ann Arbor},   \state{Michigan}, \country{USA}}}


\abstract{Women are at increased risk of bone loss during the menopausal transition; in
fact, nearly 50\% of women’s lifetime bone loss occurs during this time. In addition to level and rate of change in estradiol (E2) and follicle-stimulating hormone (FSH), the variability of these hormones may be a key predictor of bone loss; however, this remains unexplored in the existing literature. We introduce a joint model that explicitly characterizes the uncertainty in the means and variances of the hormone trajectories. Our method estimates both the individual mean marker trajectories and the individual residual variances, and links these variances to bone trajectories. In our application, for the first time, higher FSH variance interacted with time was associated with declines in bone mineral density (BMD) across the menopausal transition. At the final menstrual period, higher individual FSH variance predicted an average 0.26\% decline in BMD, but this effect is moderated over time. Our results suggest that the mean and variance of FSH, rather than E2, may be a stronger predictor of menopausal bone health. In a variety of simulation studies, our method achieves $>$ 90\% interval coverage, whereas naive two-stage alternatives often fail to propagate uncertainty in the individual-level variance estimates.}

\keywords{Bone mineral density, Estradiol, Follicle-stimulating hormone, Joint models, Subject-level variability, Variance component priors}

\maketitle 

\section{Introduction}
\label{sec:intro}
Osteoporosis, or low levels of bone mineral density (BMD), is a major public health issue as it increases the risk for fracture, a major cause of hospitalizations and morbidity \citep{marshall_meta-analysis_1996}. Although osteoporosis prevalence increases with age \citep{alswat_gender_2017}, the loss of bone mass begins during midlife (age 40-64 years) \citep{hunter_bone_2000}.  For women, declines in BMD accelerate during the menopausal transition (MT) \citep{riggs_prevention_1992, ji_primary_2015}, a period of rapid endocrinologic and physiologic changes surrounding the final menstrual period (FMP). While population level patterns of mean level and change in BMD during the midlife and late adulthood are well-established, it is equally vital to model individual trajectories of bone parameters in order to support advances in precision medicine and tailored individualized treatments. These advances in turn might alleviate individuals' anxiety and empower them with better knowledge about their health statuses and potential health trajectories during the aging process. 

For midlife women, BMD tends to gradually decline until approximately one to two years prior to the FMP. BMD decline then accelerates during this transition period, which is approximately 2 years before and after FMP \citep{sowers_design_2000,greendale_changes_2019}. After about two years post FMP, BMD stabilizes and no longer shows rapid deceleration, but still continues to decline throughout the post-menopause \citep{sowers_design_2000, greendale_changes_2019, sirola_factors_2003, recker_characterization_2000, finkelstein_bone_2008}. Previous research examined associations between estradiol (E2) and BMD during the menopausal transition, and established a general positive association between mean E2 levels and BMD \citep{sowers_design_2000,ebeling_bone_1996, park_bone_2021, tian_association_2025}. However, most of these studies have either been cross-sectional or over a relatively short time period, thereby limiting the ability to examine individual-level changes in E2. These studies also mostly focused on population level mean associations between hormones and bone outcomes. The association between \textit{individual} E2 trajectories, and in particular, individual-level E2 variability and BMD changes in peri- and post-menopausal women over a longer time period has not yet been studied.

 In addition to E2, there are characteristic changes in follicle-stimulating hormone (FSH) during the menopausal transition, with increases beginning approximately five years before the FMP \citep{randolph_change_2011}. There is an ongoing discourse in the literature as to which sex hormone – E2 or FSH – drives changes in bone across the menopausal transition. \cite{chin_relationship_2018} examined the longitudinal relationship between FSH and BMD in peri-menopausal women and found that ``rate of bone loss was inversely associated with FSH level in all subjects, regardless of BMD value". More recently, \cite{lu_associations_2025} used Spearman correlation analysis to find an association between FSH and risk of osteoporosis in menopausal and post-menopausal women.  However, \cite{gourlay_follicle-stimulating_2011} did not find a statistically significant relationship between high baseline FSH and decreases in BMD in postmenopausal younger women. These findings suggest that FSH levels may have the strongest association with bone loss during the period before the menopausal transition. 
Similar to the literature on E2, the FSH literature has largely been focused on population-level trends and associations by utilizing mean levels of FSH as predictors. These analyses have also been mostly performed on cross-sectional data \citep{lu_associations_2025} or relatively limited longitudinal datasets. However, both E2 and FSH can fluctuate substantially within \textit{individual women}, and this fluctuation has been shown to be significantly associated with various health outcomes \citep{harlow_analysis_2000, uhler_high_2005}. It would thus be of interest to understand how intra-individual variabilities, as well as overall mean hormone levels, affect women’s bone health outcomes. Furthermore, the individual variances themselves may change over time, which could also influence or predict declines in bone density. 

Our paper makes the following contributions to the existing literature on women's hormones and bone health during the midlife. Firstly, we estimate how individual hormone and bone trajectories evolve together over time, and how the relationship between these two may change with respect to time, particularly as women undergo menopause. By using a joint modeling approach, we are able to use estimated individual random effects from the hormone marker as predictors for estimating bone changes. Secondly, we estimate the individual residual variances from the hormone trajectories in order to explicitly estimate the association between hormone variances and bone health. We hypothesize that higher variability of hormones could be an indicator of broader reproductive systemic dysregulation. This in turn might impact the body's ability to maintain or preserve BMD. 

\subsection{Study of Women's Health Across the Nation Dataset}
\label{sec::swan_data}
\paragraph{Study Population} The Study of Women’s Health Across the Nation (SWAN) is an ongoing multi-site longitudinal cohort study of the menopausal transition and aging. At study baseline in 1996, participants were age 42-52 years, with an intact uterus and at least one ovary, had menses in the previous 3 months and were free of exogenous hormone use in the previous 3 months. By design, SWAN includes a sample that is racially/ethnically diverse, including White, Black, Hispanic, Chinese, and Japanese women. Since it began in 1996, there have been 17 near-annual study visits. For detailed descriptions of the measurement timings and cohort characteristics, we refer the reader to \cite{sowers_design_2000}.

\begin{figure}[h!]
\centering
\begin{subfigure}[b]{0.8\textwidth}
  \centering
\includegraphics[width=1\linewidth]{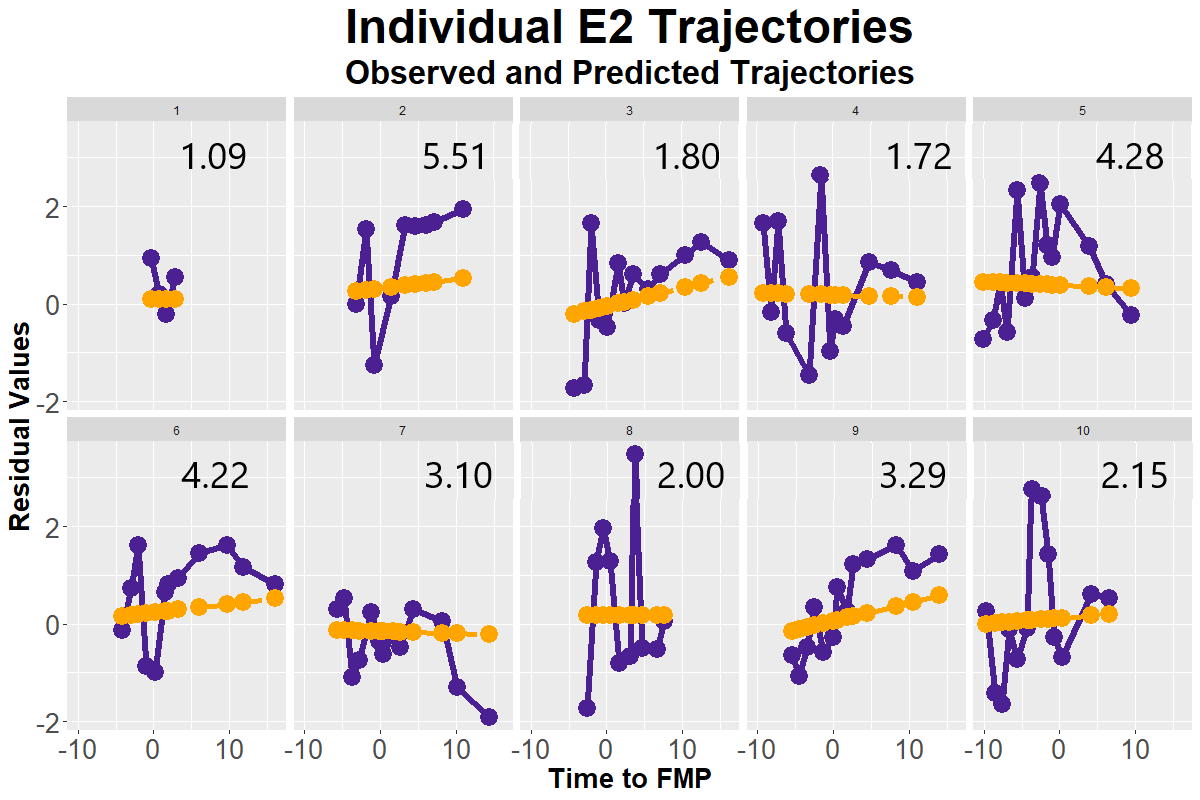}
\end{subfigure}%
\\
\begin{subfigure}[b]{0.8\textwidth}
  \centering
\includegraphics[width=1\linewidth]{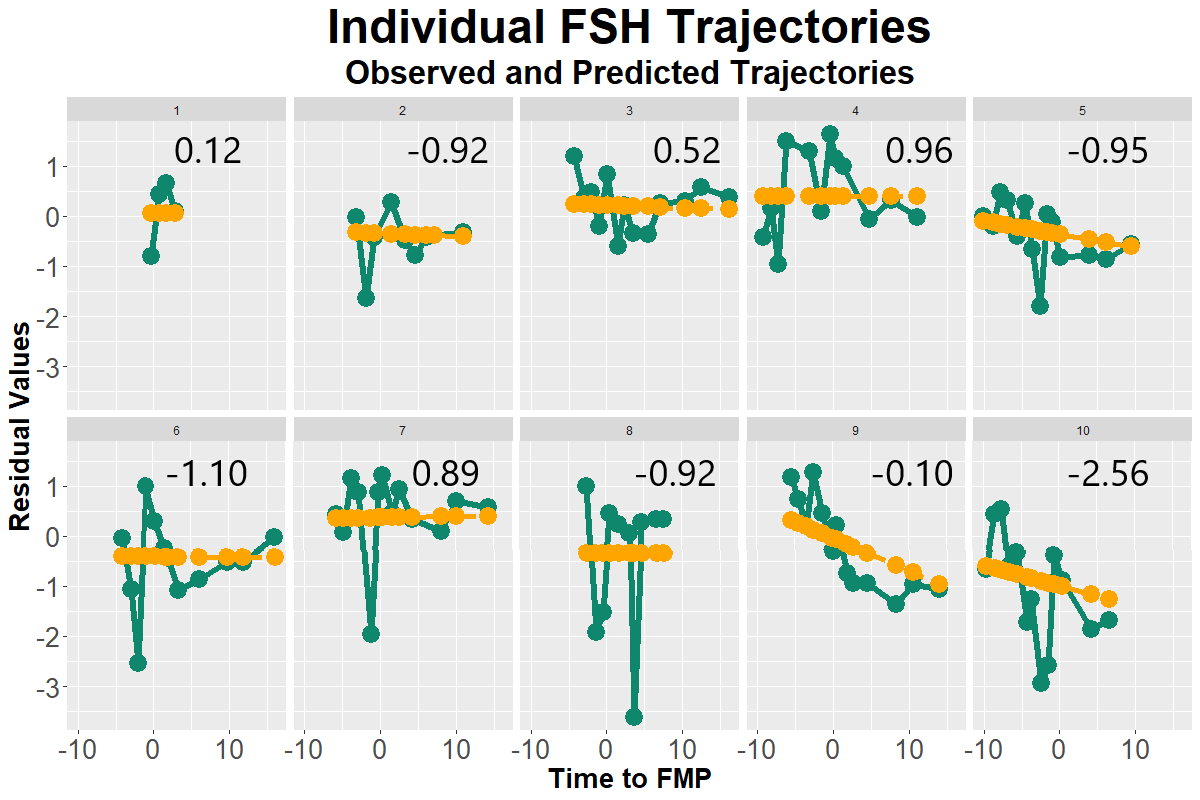}
\end{subfigure}%
\caption{Individual trajectories from 10 women in the SWAN dataset. The observed E2 (purple) and observed FSH (green) residuals have been plotted, along with their respective predicted values from fitting a linear mixed model with time to FMP as a fixed and random effect (linear slope). The sum of the model residuals for each woman are also displayed in bold text on each plot. The differences between the predicted lines and the observed residuals across individuals may indicate that specifying an individual variance term may better capture the residual \textit{individual} variances in the hormone trajectories.} \label{fig:e2_fsh_residuals_pred}
\end{figure}

At baseline, the original SWAN cohort included 3,302 women. Only five of the seven clinical sites participated in the bone protocol; thus the SWAN Bone Cohort included 2,365 women. The analytic sample included SWAN Bone Cohort women with observed sex hormone, observed FMP, and observed DEXA data across a minimum of 2 visits; 912 women were excluded for either having missing hormone values, unobserved (missing) FMP, or only baseline visits. Additional exclusion criteria included women who took hormone therapy (HRT) during their visits (n=479), since these medications can suppress BMD decline even after stopping HRT \citep{gambacciani_hormone_2014}. The final analytic dataset comprised 974 women with a total of 8,383 observations. Figure \ref{fig:e2_fsh_residuals_pred} shows the individual E2 and FSH trajectories for 10 women in our dataset. 

\paragraph{Measures}
E2 (pg/mL) and FSH (mIU/m) measurements were assayed from serum collected at baseline and during annual follow-up visits. At all visits, blood was collected fasted and during the follicular phase of the menstrual cycle (days 2-5). Details of the blood collection protocol and laboratory methodology for E2 and FSH have been published (Randolph 2011).
At each study visit, participants underwent a DEXA (dual-energy X-ray absorptiometry) scan with a Hologic densitomte (Hologic, Inc., Waltham, Massachusetts) to assess BMD of the femoral neck. Across the clinical sites, a calibration protocol for the densitometers was established and ongoing quality control measures were undertaken \citep{greendale_changes_2019}. Covariates included body mass index (BMI) and chronological age. BMI was calculated as measured weight (in kilograms) divided by measured height (in meters) squared. Chronological age was calculated as date of visit minus date of birth.

All participants provided written informed consent at each study visit and the protocols at each site were approved by each site’s Institutional Review Board.

 \subsection{Statistical Models for Longitudinal Outcomes}
 Models for longitudinal outcomes allow researchers to understand relationships between variables across time, and potentially how these associations change across time. These methods tend to fall into two broad classes: generalized estimating equations \citep{liang_longitudinal_1986} methods and mixed-effects models or structural equation modeling (SEM)-type growth curve models. While the marginal model (e.g. GEE) approach has several advantages, such as robustness in the case of misspecified marginal correlations, it is not designed for studying individual-level random effects. Mixed-effects models have been studied extensively \citep{laird_random-effects_1982, greene_fixed_2005, diggle_2013} and are well-suited for fitting multilevel or hierarchical data. These models can also easily handle interactions with time, which can be specified as an additional covariate.  
 Latent growth curve (LGC) models on the other hand provide a framework for estimating two or more trajectories simultaneously. The random effects from the LGC framework can be shown to be mathematically equivalent to the random effects specification in mixed effects models \citep{zhang_model_2022}. Generally, LGC models are not well-suited for datasets with moderate to large number of observations per individuals, since the LGC framework requires each visit or timepoint to be a unique predictor. Mixed-effects models, on the other hand, allow the time variable to be univariate and can more easily handle many observations per individual \citep{mcneish_differentiating_2018}. 

 However, both of these approaches treat the predictor variable as fixed or given, rather than being estimated as part of the model, meaning that individual level trends in the predictors are not estimated within either framework. In our particular setting, this means that we are not able to  investigate the association between individual level hormone means and variabilities with bone density, which is our primary interest.  Our modeling approach, therefore, needs to explicitly estimate these individual level quantities, which are unobserved \citep{ carroll_measurement_2006, elliott_associations_2012, chen_variance_2024}. Extending the mixed effects model to the joint modeling framework, where we can simultaneously estimate the longitudinal predictor and longitudinal outcome \citep{olsen_two-part_2001,lo_joint_2017}, is thus a compelling approach for our research setting. 
 
 Another noticeable gap in both frameworks is that the individual residual variances in the predictors are usually treated as nuisance parameters, rather than as potentially important entities for predicting the outcome. There has been a growing collection of research \citep{elliott_associations_2012,jiang_joint_2015, gao_comparing_2022} on joint models that evaluate the associations between within subject variability and outcomes of interest; however, these models have focused on cross-sectional outcomes. Methods for estimating individual variances from longitudinal markers to predict longitudinal outcomes are currently lacking in the literature. Our proposed joint model contributes to this research by providing a general method for estimating individual-level variability from a longitudinal marker and using this variability to predict a longitudinal time-varying outcome. By specifying a joint model for the predictor and outcome, we are able to estimate these two trajectories simultaneously by linking these via shared individual random effects. Our specification can easily extend to incorporating a time-varying variance parameterization, as demonstrated by simulation studies. 

The rest of this paper is organized as follows. We describe our Bayesian joint model framework in Section \ref{sec::model_form}. Section \ref{sec:Simulation} introduces simulation studies that demonstrate that our model produces less biased estimates of the outcome regression coefficients than alternative two-stage models that do not account for the statistical uncertainty  in the individual means and variances. We apply our model in Section \ref{sec::swan_application} to the SWAN dataset, where we focus on the study of the associations between E2 and FSH mean trajectories and variances and BMD outcomes in midlife women. We believe that this is the first scientific assessment in using individual level variances of E2 and FSH to predict declines in longitudinal bone measures. Finally, in Section \ref{sec::discussion}, we discuss the implications of our findings along with future directions. Our supplementary material contains further details about the model validation for the data analysis (e.g. posterior predictive checks). We provide R code to run our model via an open-access GitHub repository (see supporting information at the end of this article).

\section{Joint Model}
\label{sec::model_form}
In this section, we describe our proposed model, which specifies our longitudinal predictor $X_{ij}$, which is measured at each timepoint $t_{ij},j=1, \ldots, n_{i}$, for each subject $i = 1, \ldots, N$ where $n_i$ is the total number of observations for individual $i$. Our joint model describes how $X_{ij}$ and $Y_{ij}$ simultaneously evolve over time, and how these two trajectories are linked via individual-specific vectors of regression coefficients $\bb_{i}$ and residual variances $\sigma_{ij}$.  
\subsection{Proposed Model}
\label{sec::model_likelihood}

\subsubsection{Longitudinal Marker Model}
\label{sec:::predictor_model}The model for the individual predictor trajectories is given by: 
\begin{align} 
& X_{ij} \mid \bb_i, \sigma_{ij}^{2}, t_{ij} \sim  \mathcal{N}\left(\mu(t_{ij}; \bb_i), \sigma_{ij}^{2}\right),     j=1, \ldots, n_{i}, {\sf ~independently~for~} i=1, \ldots, N,  \label{eq::marker_model} \\
 & \bb_{i} \overset{\sf indep.}{\sim}  \mathcal{N}_P(\balpha, \Sigma), {\sf ~independently~for~} i=1, \ldots, N,
   \label{eq::individual_beta_sigma}
\end{align}
where $\mathcal{N}(\mu,s)$ represents a Gaussian distribution with mean $\mu$ and variance parameter $s$ and $\mathcal{N}_{P}(\balpha,\Sigma)$ is the $P$-dimensional generalization the Gaussian distribution with mean vector $\balpha$ and variance-covariance matrix $\Sigma$. $X_{ij}$ has two components: the mean $\mu(t; \bb_i)$, which is a function of time and a vector of $P$ regression coefficients, $\bb_{i}=(b_{i1}, \ldots, b_{iP})^{\transp}$, and $\sigma_i^{2}$, the individual-level time-invariant residual variances. The dimension of the regression coefficients, $P$, is the number of basis functions of time, and is pre-specified in advance. In this paper, we consider linear forms for $\mu(\cdot)$ (e.g. individual intercepts and slopes, $P =2$). This is reasonable given the limited number of observations per woman combined with the preliminary population detrending (see Section \ref{sec::swan_application} for details), and it has the added advantage of providing easy interpretation for the outcome model coefficients. However, in general, non-linear or semi-parametric extensions of $\mu(\cdot)$ may also be considered for the predictor model \citep{rizopoulos_bayesian_2011,li_flexible_2021}. We note that the increased flexibility of a higher order mean function may come with computational issues during the estimation process. The individual-level residual variance, $\sigma_{ij}$, is a latent random variable that may depend on $t_{ij}$; in the case of time-invariant variances, this can simplify to $\sigma_i$. 

\paragraph{Priors for \textbf{\(\balpha\)}:}
We assume that the population mean regression coefficients, $\balpha = (\alpha_{1}, \ldots, \alpha_{P})^\transp$, are drawn from a $P$-dimensional Normal distribution: 
\begin{align}
 &  \balpha ~{\sim}~ \mathcal{N}_P(0, \xi_0^2I_{P\times P}), \label{eq::prior_Bi} \\
  & \Sigma  =  K L K, ~K  =\diag\{k_{1}, \ldots, k_{P}\}, \label{eq::prior_Sigma} \\ 
  & k_{p} \sim \textsf{half-Cauchy}(0, \tau_0), p = 1, \ldots, P, L \sim {\sf LKJ}(\zeta),
   \label{eq::hyperprior_Sigma}
\end{align} 
where $K =\diag\{k_{1}, \ldots, k_{P}\}$ is a diagonal matrix and $L$ is a correlation matrix, with a Lewandowski-Kurowicka-Joe (LKJ) diffuse prior \citep{lewandowski_generating_2009}. The values of $\xi_0, \tau_0, \zeta$ parameters are set a priori. In our application, we also do not have strong \textit{a priori} information about these parameters; hence our choice of weakly informative priors.  
\paragraph{Time-varying $\sigma_{ij}^{2}$:} A model that allows for the individual variances, in addition to the means, to change over time can be specified with the following parameterization: 
\begin{align}
   &\log(\sigma_{ij}^{2}) =  g(t_{ij}; \blambda_i), \\
  & \blambda_i \sim \mathcal{N}(\bdelta, \Psi),\bdelta \sim \mathcal{N}(0, \xi^{2}_\delta I_{P_{S} \times P_{S}}), \Psi \sim {\sf LKJ}(\zeta_{P_{S}}), 
  \label{prior_sij}
\end{align}
where $\blambda_i$ is a $P_S$-dimensional vector that represents the individual-level variance regression coefficients, with $P_S$ being specified in advanced. As in Equations \ref{eq::prior_Bi} - \ref{eq::hyperprior_Sigma}, the hyperparameters for $\bdelta, \Psi$ would also be set a priori. In general, since estimating individual variances will be more computationally intensive, we recommend setting $P_S \leq P$; that is, allowing more flexibility in the mean regression function, unless there is reason to believe that a more flexible variance trend is justified. It is also critical that there be a sufficient number of longitudinal observations within a subject to detect differences in variability over time.

Simulation study \ref{sec:sim2} evaluates the relative advantages of this time-varying variance over the time-invariant one under different data generating scenarios. In practice, when applying our model to the SWAN dataset, we found that the time-varying variance model had severe convergence issues compared to the time-invariant variance model. We discuss this further in Section \ref{sec::swan_application}.  

\paragraph{Prior for \(\sigma_{i}^{2}\)} In the case of time-invariant individual variances, the $\sigma_i^{2}$'s are drawn from a log-Normal distribution, i.e.: 

\begin{align} 
  &  \log(\sigma_{i}^{2})\sim \mathcal{N}(\nu, \psi^2),   \label{eq:prior_sigmai} \\
&  \nu \sim \mathcal{N}(m, \xi^2), \psi \sim \textsf{half-Cauchy}(0,\tau), 
 \label{eq:hyperprior_sigmai}
 \end{align} 
 
where the values of hyperprior parameters $m, \xi, \tau$ are set a priori. 
Following the recommendations made in \cite{gelman_prior_2006}, we use the half-Cauchy hyperprior on $\psi$, the square root of the variance parameter. Although the inverse-Gamma distribution is the conjugate prior for the variance parameter of a Gaussian distribution, inferences using the inverse-Gamma distribution can be extremely sensitive to the choice of hyperparameter values \citep{gelman_prior_2006}. The half-Cauchy distribution avoids this potential issue due to its heavier tail, which still allows for higher variance values.

\subsubsection{Longitudinal Outcome Model}
\label{sec:::outcome_model}
The outcome variable, $Y_{ij}$, is related to individual-specific mean and variance parameters $\bb_i$ and $\sigma_{ij}^{2}$ (Equation \ref{eq::marker_model}) via the following specification: 

\begin{align}
 \label{eq::outcome_regression}
  Y_{ij} \mid \bb_i,\ba_i,t_{ij}, \sigma_{ij}^{2}, \Wb_{ij} \sim &  \mathcal{N}\left(\eta_{ij}(\mu_{ij}, t_{ij}, \sigma_{ij}^2, \Wb_{ij}, \ba_{i}),  \sigma_{Y}^2\right),
  \\ i=1, \ldots, N, j=1, \ldots, n_{i}, \nonumber
\end{align}
{where $\mu_{ij}$ is the mean from the marker model, $\bW_{ij}$ are additional covariates that may be associated with the outcome, $Y_{ij}$, $\ba_i$ are the individual random effects for $Y_{ij}$, and $\sigma_{Y}^2$ is the residual variance for $Y_{ij}$. Here, $\mu_{ij}$ ($\bb_i$) and $\sigma_{ij}^{2}$ are linked to the longitudinal outcome in the mean regression function $\eta_{ij}$.
Using $\mu_{ij}$ in the outcome mean model allows us to evaluate the direct relationship between the model-estimated predictor value and the outcome $Y_{ij}$, rather than linking $X_{ij}$ and $Y_{ij}$ via the mean coefficients $\bb_i$. This ease of interpretation may be preferred in scientific applications. As an example, the form for $\eta_{ij}$ in  Section \ref{sec::model_application} is a multiple linear regression with interactions between 1) the predicted mean marker values, $\mu_{ij}$, and time and 2) the estimated residual variance $\sigma_i$ and time. This formulation of $\mu_{ij}$ also avoids contamination from the time-invariant $\bb_i$. 

 The individual residual marker variances, $\sigma_{ij}^{2}$, could also be specified similarly if we allow these to vary over time (see simulation study \ref{sec:sim2}). We write the additional covariates $\Wb_{ij}$ as time-varying, but these of course, can be time-invariant (in our application, we adjusted for starting BMI and starting age of each participant, which are constant across time). In our application, we focus on simple specifications of $\eta(\cdot)$, e.g., linear models with two-way interactions in order to maintain interpretability of the coefficients. However, extensions with more complex interactions or non-linear mean structures are also possible, with the caveat that the interpretability of the coefficients may be challenging with higher-order terms.

\paragraph{Priors for $\bbeta$, $\textbf{a}_i$, $\sigma^{2}_{Y}$}
\label{prior_outcome_model}
\par For the outcome model, we use independent $\mathcal{N}(0, 10^2)$ priors for each element of the outcome regression parameters ($\bbeta$, $\bbeta^W$), and a diffuse prior on the outcome residual standard deviation parameter 
$\sigma_{Y} \sim \textsf{half-Cauchy}(0,2.5)$, as recommended by \cite{carpenter_stan_2017}. For $\ba_i$, the random effects, we place a multivariate Gaussian prior with mean zero and precision matrix $\tau_a$, i.e., $\ba_i \sim \mathcal{N}(0, \tau_a)$. In the case of a random intercept $a_i$, $\tau_a$ can be drawn from a half-Cauchy distribution or, in the case of a vector-valued $\ba_i$, $\tau_a$ is a covariance matrix, whose values can be drawn from the prior described in Equation \ref{eq::prior_Bi}.

\paragraph{Joint Distribution} Let $D= (Y_{ij}, X_{ij}, t_{ij}, \Wb_{ij})$ denote the observed data, $Z$ = ($\bb_i$, $\sigma^{2}_{ij}$, $\ba_i$) denote the subject-level latent variables, and $\Theta = (\balpha, \Sigma, \blambda, \Psi, \bbeta, \bbeta^{W}, \tau_a, \sigma^{2}_{Y})$ denote the model parameters. 
We also let $\pi(\Theta)$ denote the prior distribution of the parameters in $\Theta$: 
\begin{align}
\pi(\Theta)= \pi(\balpha) \pi(\Sigma) \pi(\xi)  \pi(\blambda)\pi(\Psi) \pi(\bbeta, \bbeta^{W}) \pi(\sigma^{2}_{Y}).
\end{align}
Additionally, if we assume 1) the prior distributions for each parameter in $\Theta$ as given in Sections \ref{sec:::predictor_model} and \ref{sec:::outcome_model}, and 2) each parameter in $\Theta$ has an independent prior, we can then write the joint distribution of $D$, $Z$, and $\Theta$ as:
\begin{align}
   P(\Theta, D, Z)  \propto &  \prod^{N}_{i=1}  \Big\{ \frac{1}{\sqrt{2\pi|\Sigma|}}{\exp}(-\frac{1}{2}(\bb_{i}-\balpha)^{\transp} \Sigma^{-1}(\bb_{i}-\balpha)) \nonumber
    \\ 
    & \times \frac{1}{\sqrt{2\pi |\Psi|}}
    {\exp}( -\frac{1}{2} \log(\sigma_{ij}^2)-g(t_{ij}; \blambda_i))^{\transp} \Psi^{-1} (\log(\sigma_{ij}^2)-g(t_{ij}; \blambda_i))
    \nonumber \\ 
    & \times \frac{1}{\sqrt{2\pi|\tau_a|}}{\exp}(-\frac{1}{2}\ba_{i}^{\transp}\tau_{a}^{-1}\ba_i) \Big\} \nonumber \\ 
     & \times \prod ^{N}_{i=1}  \prod ^{n_i}_{j=1}  \Big\{\frac{1}{\sqrt{2\pi \sigma_{ij}^{2}}}{\exp}\left(-\frac{1}{2}\{\dfrac{X_{ij}-\mu_{ij}(t_{ij}; \bb_i)}{\sigma_{ij}}\}^{2}\right)  \nonumber 
     \\
    & \times \dfrac{1}{\sqrt{2\pi\sigma_{Y}^{2}}}  {\exp} \left(-\frac{1}{2}\frac{(y_{ij}-\eta_{ij}(\bb_i, \sigma_{ij}^{2}, \Wb_{ij}, \ba_i; \bbeta,\bbeta^W))^2}{\sigma_{Y}^{2}}\right) \Big\} \nonumber 
    \\&  \times \pi(\Theta). \label{posterior_likelihood}
\end{align}

\subsection{Posterior Inference}

We implemented our joint model using Stan and the \verb"rstan" package \citep{rstan} to obtain the posterior estimates. Since the posterior distribution for our joint model is intractable, we used MCMC sampling via the HMC (Hamiltonian Monte Carlo) algorithm implemented by Stan \citep{carpenter_stan_2017}. Stan is a probabilistic programming language that provides a self-tuning, optimized implementation of HMC. 

For our two simulations studies in Sections \ref{sec:sim1} and \ref{sec:sim2} we ran two chains per independent replicate data set, with $2,000$ iterations with $1,000$ burn-in. For the E2 predictor model in Section \ref{sec::swan_application}, we ran 2 chains each with 4,000 iterations and 2,000 burn in. For the FSH model, we found that running 2 chains with 2,000 iterations with 1,000 burn in was sufficient for the model to achieve convergence. We conducted visual inspection of the traceplots for all model parameters which indicated non-divergent chains. All of the chains in each model were combined for computing posterior summaries.

We also used Stan's R-hat convergence diagnostic \citep{vehtari_rank-normalization_2021} to evaluate model convergence. The R-hat estimate compares the between-chain estimates with the within-chain estimates produced by the model \citep{brooks_general_1998}. High R-hat values can indicate that the model has not converged and that the chains have not mixed well. In our application, the coefficients related to the individual random effects (hormone means and variances) all had R-hat values $<$ 1.05, indicating that the chains had converged. Additionally, model checks of the posterior predictive distribution in the supplementary materials (Section S2) indicated that our models generated reasonable predictions for our datasets.

\section{Simulation Study}
\label{sec:Simulation}
The goal of our simulation studies was to 1) evaluate our model's operating characteristics and 2) compare against common alternatives that could also be used in modeling individual means and variances as predictors of longitudinal outcomes.
For each proposed method and each parameter $\theta$, we assessed the 1) bias (defined as $\dfrac{1}{R} \sum_{r=1}^{R}(\hat{\theta}^{(r)} - \theta_0)$ where $\hat{\theta}^{(r)}$ is the posterior mean of $\theta$ obtained from the $r$-th replication), 2) the coverage rate of the nominal $95\%$ credible intervals (CrI; defined as $\dfrac{1}{R}  \sum^{R}_{r=1} \mathbbm{1}\{\theta_0 \in I_r\} $ where 
$I_r$ is the $95\%$ CrI for parameter $\theta$ obtained by computing the 2.5\% and 97.5\% percentiles of the draws from the posterior distribution for the $r$-th replication, and 3) average length of the 95\% CrIs obtained across simulation replicates, defined as $\dfrac{1}{R}  \sum^{R}_{r=1} T_{r} $, where $T_{r}$ is the length of $I_r$, i.e., the range of the estimated 2.5\% and 97.5\% posterior quantiles for $\theta$ in replicate $r$. The following sections present simulation results for a time-invariant individual variance model (Section} \ref{sec:sim1}) and time-varying individual variances (Section \ref{sec:sim2}). We also provide additional simulation results where we varied the number of individuals and the number of observations per individual to evaluate how the model performs under these different data-generating scenarios (Supplementary Material S4).

\subsection{Simulation I: Constant Individual Variance Model}
\label{sec:sim1}
\begin{figure}[!t]%
    \begin{subfloat}
        \centering
        \includegraphics[scale=0.12]{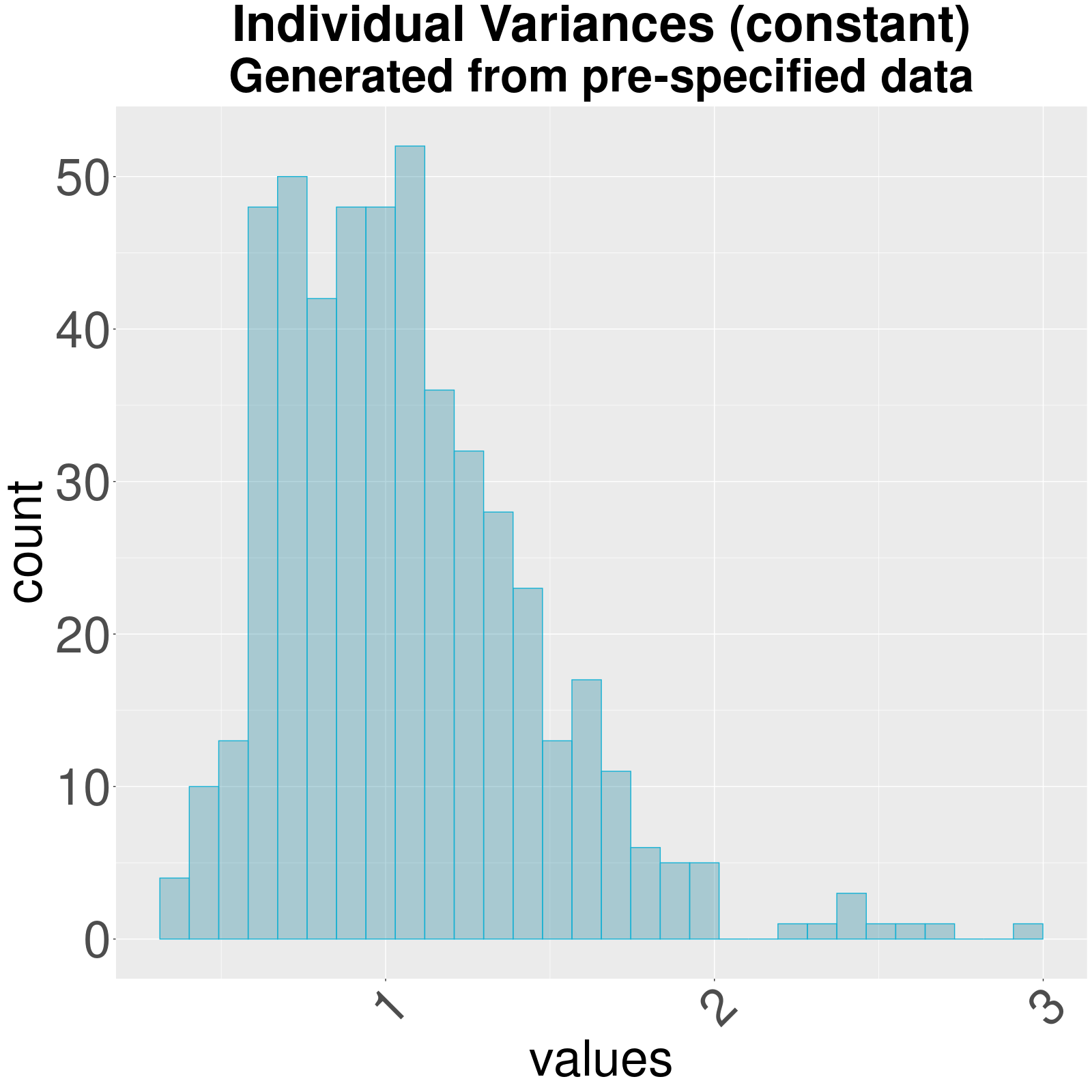}
    \end{subfloat}
\begin{subfloat}
   \centering
\includegraphics[scale=0.12]{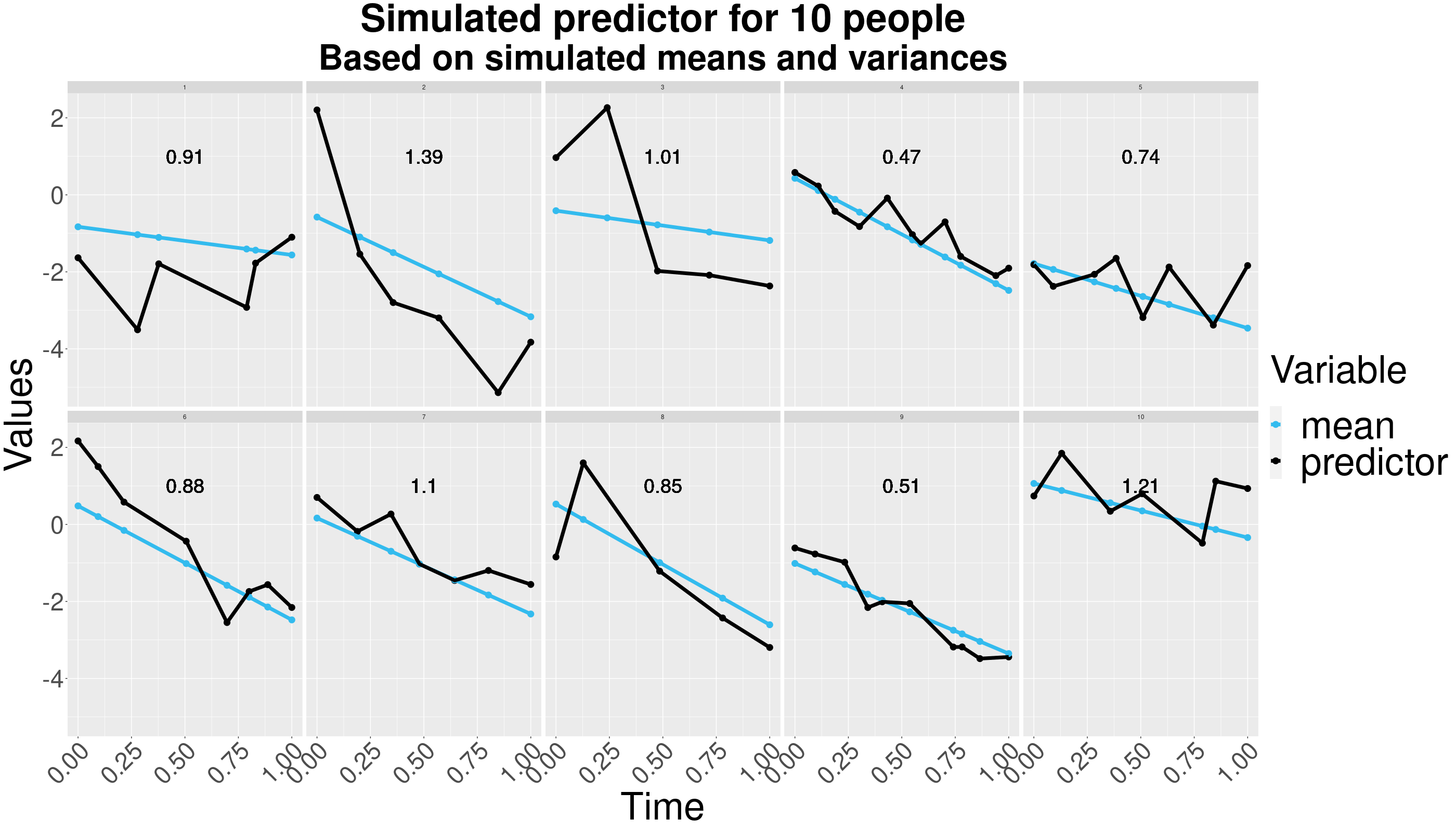}
    \end{subfloat}
    \caption{Histogram of individual variances (left plot) and individual mean and marker trajectories (line plots) for 10 individuals, along with the generated individual variance labelled in each plot.}
        \label{fig:tvv_sim1_constant_plots}
\end{figure}

For this simulation study, we generated data for N = 300 individuals. We simulated between 2 to 15 timepoints for each individual, which mimics the SWAN study dataset. Based on these individual timepoints, we then simulated the marker values for each individual using the following data generating parameters: 
\begin{align*}
& X_{ij}  \sim  \mathcal{N}(\bm{\mu}_{ij}, \sigma_i^{2}),  
\bm{\mu}_{ij} = b_{i0} + b_{i1} t_{ij}, \\
& \bb_{i} \sim \mathcal{N}_2\left(\balpha, \Sigma \right),  \balpha =(0, 2)^\transp, \\
&\Sigma =\begin{pmatrix}
1 & -0.05 \\
-0.05 & 1 
\end{pmatrix},  \log(\sigma^{2}_{i}) \sim \mathcal{N}(0, 0.375^{2}).
\end{align*}
Figure \ref{fig:tvv_sim1_constant_plots} displays the simulated marker means, marker trajectories, and constant variances that are generated by these specified parameters.

For the longitudinal outcome, we assumed the following model: $Y_{ij} \sim \mathcal{N}(\eta_{ij}(\bb_i, \sigma_i^{2}, a_i, t_{ij}), \sigma_{0}^2)$ and set 
\begin{align*}
  \eta_{ij}(\bb_i, s_i) = \beta_{0} + \beta_{1} \mu_{ij} +\beta_{2} \sigma_i^{2}  \\ + (\beta_{3} + \beta_{4} \mu_{ij}  +\beta_{5} \sigma_i^{2}) t_{ij} + a_{i},
\end{align*}
where the true values of $\bbeta$ are shown in Table  \ref{tab::sim1_outcome_betas}. We specified the random intercept $a_i$ as follows: 
\begin{align*}
    a_i \sim  \mathcal{N}(0, 0.5^{2}).
\end{align*}
Lastly, we set $\sigma_{0} = 0.1$. In this simulation, we did not adjust for other covariates $\Wb_i$ in either submodel. We present the results for 200 replicates in Table \ref{tab::sim1_outcome_betas} for the outcome submodel parameters $\bbeta$.

\subsubsection{Alternative Methods}
\label{sec:alternatives}
We compared our approach against two alternative two-stage methods: a two-stage linear mixed model (TSLMM), and a two-stage Bayesian model with longitudinal outcome (TSLO). The TSLO approach is essentially the two-stage version of our joint model, where we used the posterior mean estimates from the means and variances of the longitudinal marker to predict the outcome in the second stage. We refer to our joint model as the ``Jointly Estimated (Model) with Longitudinal Outcome", or JELO. In the absence of a joint model, these approaches would be reasonable methods for scientific researchers who wish to analyze the associations between a longitudinal predictor and longitudinal outcome. However, as shown in previous literature, two-stage methods often do not correctly preserve the uncertainty associated with estimating the individual random effects from the predictor marker variable \citep{hickey_joint_2016}. 
\label{subsec::other_methods}

\paragraph{Two-Stage Linear Mixed Models (TSLMM)}

In the first stage, we fit a linear mixed model with the following specification:  
\[X_{ij} = \beta_{0} +  b_{i0} + \beta_{1} t_{ij} + b_{i1} t_{ij}  + \epsilon_{ij}. \]

We used an unstructured variance-covariance structure for the random effects, which is the default specification for the \textbf{{lmer}} package. The \textbf{{nlme}} package yielded very similar results. 

 We obtained the predicted values of $x_{ij}$ using the \verb|predict()| function.  We estimated $\sigma_i^{2}$ by first computing the model residuals (e.g. $X_{ij} - (\hat{B}_{i0}+ \hat{B}_{i1}t_{ij}$)), where these $``B_{i}"$ coefficients are defined as $\hat{B}_{i0}$ =  $\hat{\beta}_{0} + \hat{b}_{i0}$ and $\hat{B}_{i1}$ = $\hat{\beta}_{1} + \hat{b}_{i1}$,  where $\hat{\beta}_{pi}$ and $\hat{b}_{pi}, p =0, 1$, and then computed the variance across all residuals.

In the second stage model, again using the \textbf{{lme4}} package in R, we fit another linear mixed model with the following specification:
\[Y_{ij} = \beta_{0} + \beta_{1} \hat{x}_{ij} + \beta_{2} t_{ij} + \beta_{3} \hat{\sigma}_{i}^{2} + \beta_{4}  \hat{x}_{ij}t_{ij}  + \beta_{5} \hat{\sigma}_{i}^{2} t_{ij} + \epsilon_{ij}. \]

\paragraph{Two-Stage Individual Variances (TSLO) Model} We used Equations (\ref{eq::marker_model}) and (\ref{eq::individual_beta_sigma}), and the prior specifications in Equation (\ref{eq::prior_Bi}) to (\ref{eq:hyperprior_sigmai}) to fit the longitudinal predictor model. We then collected the posterior mean estimates of $X_{ij}$  and $s_i^{2}$. To obtain the posterior mean of $X_{ij}$, we computed the mean across all $X_{ij} \sim \mathcal{N}(b_{i0} + b_{i1} t_{ij}, \sigma_{i}^{2})$ drawn at each MCMC iteration. The posterior mean of $\sigma^{2}_{i}$ is obtained similarly. Once we computed the posterior means of $X_{ij}, \sigma_{i}$, we used these values in the outcome model (Equation \ref{eq::outcome_regression}), along with the prior specifications on the outcome regression coefficients in \ref{prior_outcome_model}).

\subsubsection{Simulation I: Results}

Table \ref{tab::sim1_outcome_betas} presents the results of Simulation I. We see that for $\beta_{3}$ and $\beta_{5}$, the coefficients of the variance parameters in the outcome submodel, the biases from the TSLMM approach and the TSLO approach are higher than the bias from our proposed model. Additionally, the coverage of the true parameters is extremely low, with neither alternative being able to achieve $>50\%$ coverage. This indicates that if the variability of the longitudinal predictor is indeed important for estimating the outcome, neither two-stage alternative would be able to consistently estimate this association. 

\par In particular, we see that our model outperforms the two competitors with the regards to estimating the variance coefficients ($\beta_3, \beta_5$). The TSLMM has extremely low coverage, which makes sense because this model framework does not account for individual variability. The two stage approach, TSLO, performs somewhat better, but fails to achieve  $>50\%$ coverage for either parameter. Our joint modeling framework explicitly models the individual level variances and thus appropriately carries over the uncertainty from the variances into the second submodel, which improves estimation of the parameters in the outcome regression.

\subsection{Simulation Study II: Comparison of constant variance and time varying variance}
\label{sec:sim2}

There were two main objectives of this study. The first was to understand how well our model could recover the data generating parameters with a time-varying individual variance component.  The second objective was to compare this approach to the approach with the time-invariant individual variance. This comparison gave us more insight into the situations where not specifying the time-varying component could result in large biases or high undercoverage of the true parameters. 

We evaluated two scenarios with time-varying individual variances. For each simulation replicate, we generated data for N = 500 individuals and gave each individual between 4 to 12 timepoints.  

In the first scenario, we simulated the marker values for each individual using the following parameters: 
\begin{align*}
& X_{ij}  \sim  \mathcal{N}(\mu_{ij}, \sigma_{ij}^{2}),
\bm{\mu}_{ij}=  b_{i0} + b_{i`} t_{ij}, \\ 
& \bb_{i} \sim \mathcal{N}_2\left(\balpha, \Sigma \right), 
\balpha =(0, -2)^\transp,  \Sigma =\begin{pmatrix}
1 & -0.25 \\
-0.25 & 0.5 
\end{pmatrix}, \\
& log(\sigma^{2}_{ij})  = \lambda_{0i} + \lambda_{1i} t_{ij}, \blambda_i \sim \mathcal{N}_{2}(\bdelta, \Psi), \\
& \bdelta =(-1, 0.5)^{\transp},  \Psi  =\begin{pmatrix}
1 & 0.1 \\
0.1& 0.5
\end{pmatrix}, 
\end{align*} 
so that the individual intercepts and slopes for the variance trends are larger in magnitude. We will refer to this scenario as the ``high-variability" (HV) case.

In the second scenario, we kept the same $\balpha, \Sigma$ values, but changed $\bdelta, \Psi$ to be: 
\begin{align*}
\bdelta =(0, 0)^{\transp},  \Psi =\begin{pmatrix}
0.5 & -0.01 \\
-0.01 & 0.05
\end{pmatrix}, 
\end{align*}
so that the intercepts and slopes for the individual variances were smaller in magnitude.  We will refer to this scenario as the ``low-variability" (LV) case. 

We assumed the same mean function for the outcome model for both cases: 
\begin{align*}
\eta_{ij}(\bb_i, \sigma^{2}_{ij}) = \beta_{0} + \beta_{1} \mu_{ij} +  \beta_{2} \sigma^{2}_{ij} \\ + (\beta_{3} + \beta_{4} \mu_{ij} +\beta_{5}  \sigma^{2}_{ij}) t_{ij} + a_{0i},
\end{align*} so that the marker means and variances are also interacted with time. The random intercepts $a_{0i}$ are drawn from a $\mathcal{N}(0, 0.5^{2})$ (as in the previous simulation study in Section \ref{sec:sim1}). Lastly, we set the value of the residual standard deviation of the outcome to be $\sigma_0 = 0.1$. We have created visualization for both variance scenarios (HV and LV): Figures 2 and 3 in the supplementary material  display histograms of individual intercepts and slopes of the variances for one simulation replicate and also 10 individual marker trajectories, based on these simulated variances and means.

\subsubsection{Model Comparisons}

We compared our model (JELO) with a time-varying variance performance against our model with a constant variance  parameterization (JELO CV). For the JELO CV simulation replicates, we used the same simulated datasets generated by the time-varying variance setup described above, but we fit the model for individual variances described in Equations (\ref{eq:prior_sigmai}) and (\ref{eq:hyperprior_sigmai}).

From Table \ref{tab::sim2_outcome_betas}, it is clear that when there are large individual time-varying variances, incorrectly assuming constant individual variances leads to undercoverage in the individual variance outcome parameters. The bias is large and the coverage of the coefficient for the individual variance $\beta_2$ and the time-variance interaction coefficient $\beta_5$ are essentially non-existent. However, if the individual variances are smaller, then the assumption of constant variances does not appear to result in undercoverage or substantially biased estimates of the true parameters. We can be reasonably confident that the constant variance assumption may suffice in scenarios where the individual variances are time-varying, but not systematically changing across time. However, if the time-varying variances are indeed changing across time within an individual, then this assumption would likely result in higher bias and substantial undercoverage of the true parameters.

\section{Associations between hormone levels and hormone variabilities, and bone trajectories}
\label{sec::swan_application}

\subsection{Data Processing}
Based upon population-level data, E2 and FSH demonstrate well-established and characteristic changes during the menopausal transition \citep{randolph_change_2004}. To remove the population trend in the hormones, we fit a loess curve to each hormone using the \verb|loess| function in R and subtracted each women's measurements from this loess fit. This detrending allowed us to focus on estimating the individual level hormones. We performed all of the following analyses in this section using these hormone residuals. Figure \ref{fig:e2_fsh_residuals_pred} displays the E2 residuals and the FSH residuals for 10 women in our dataset, along with predicted values estimated by fitting a linear mixed model, fit eparately to each set of hormone residuals. The linear mixed model was estimated via the \verb|lmer| R package. From the figures, we can see that after estimating the mean trajectories, there is still substantial difference in the individual level variability across subjects. The mean model for the hormone predictor was a linear regression with a random intercept and slope, i.e.: $\mu_{ij} = b_{i0} + b_{i1} t_{ij}$. 
 
For all analyses, the time scale was considered as time to/from the FMP. FMP was defined retrospectively following 12 months of amenorrhea. Longitudinal analyses were conducted to estimate the association between hormone variability and femoral neck BMD. We also lagged the hormone residuals by 1 visit, for ease of interpretation. Additionally, interaction terms were included and examined to determine whether the relationships change over time. Thus, our statistical objective is to develop a modeling framework that can estimate the individual level mean and variance of a longitudinal marker, and also link these estimates to the longitudinal outcome of interest via a regression function. 

\begin{figure}
    \centering
    \includegraphics[scale=0.2]{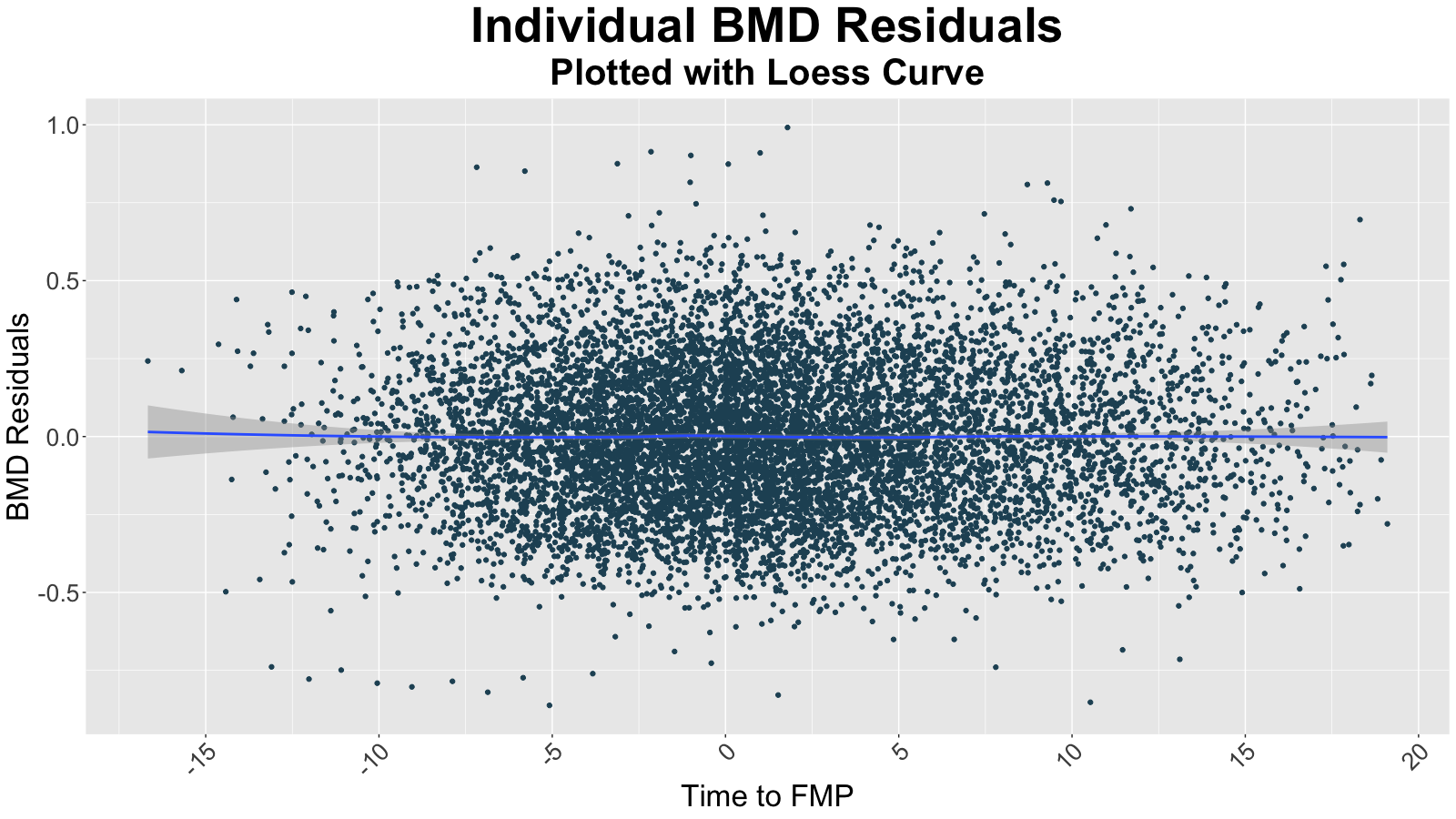}
        \caption{Plot of BMD residuals for all individuals in our dataset, plotted over time to FMP. A loess curve has been added to show the average population trend of the residuals. Prior to detrending, the BMD observations had been $\log_{2}$ transformed.}
    \label{fig::bmd_resid}
\end{figure}

A base 2 log transformation was used on the outcome of interest (BMD). In order to focus on the \textit{individual-level} changes in BMD trajectories, measurements were detrended by fitting a loess curve to all measurements (using the loess function in R) and then subtracting the individual measurements from the fitted loess values \citep{elliott_associations_2012,jiang_joint_2015}. Figure \ref{fig::bmd_resid} displays the residual BMD values after our detrending, in order to simplify subject level trends as lower-level polynomial functions; in practice a linear approximation appeared sufficient after population-level detrending.

Finally, the outcome model was adjusted for baseline BMI and baseline age. To ease interpretability, baseline BMI measurements were standardized to be relative to the population mean (baseline) BMI and baseline age was centered relative to the population mean (baseline) age. Summary statistics of all variables in the joint model are shown in Table \ref{tab:swan_descriptive_stats}.  

\subsection{Model Application to the SWAN Dataset}
\label{sec::model_application}
We now apply our proposed model to analyze the hormone and bone trajectory data described in Section \ref{sec::swan_data}. Our outcome model formulation was specified as follows: 

\begin{align*}
   \EE(\log_{2}(BMD_{ij})) =  \beta_0 + \beta_1 \mu_{ij} + \beta_2 \sigma^{2}_{i}
   + (\beta_3  + \beta_4 \mu_{ij} +  \beta_5 \sigma^{2}_{i} )t_{ij} \\ + \beta_6 BMI^{*}_{i} +  \beta_7 Age^{*}_{i} + a_{0i},
\end{align*}

\noindent
where $\mu_{ij}$ is the mean E2 (FSH) residual at time $t_{ij}$ from the longitudinal submodel, $\sigma^{2}_{i}$ is the individual-level variance, $t_{ij}$ is the time to FMP for each individual woman, $BMI^{*}_{i}, Age^{*}_{i}$ are the standardized BMI and baseline age values for each individual, and $a_{0i}$ is a random intercept for each woman. In this model formulation, the regression coefficient terms in $\eta_{ij}$, the mean function, are linearly related to the outcome. 

We ran two separate models, one with E2 measurements as the main biomarker measurement of interest and one with FSH measurements as the main predictor of interest. For the longitudinal predictor, we used the E2 (FSH) measurement obtained at the previous visit to predict BMD at the following visit. This is to better capture how differences in E2 at an earlier time may be associated with BMD declines later, rather than analyzing E2 and BMD values at the same timepoint. We checked model convergence by examining the traceplots and the R-hat values of the parameters. We have included selected traceplots for the outcome model parameters in S1 of the supplementary material . Additionally, posterior predictive checks performed on the outcome (BMD) in both models supported that our model was able to generate data that is consistent with the observed values (See Figures 1 and 2 in the supplementary material). 

We further checked if a random intercept in the outcome model was sufficient by running the models with 1) a linear random slope and 2) a linear and a quadratic random slope for the outcome (BMD), the estimates of the regression coefficients did not change substantially (in magnitude or direction). We also found that the posterior mean estimates of the variances of the linear random slopes and the quadratic random slopes were essentially 0. Furthermore, the E2-BMD model with a quadratic random slope failed to converge. We thus concluded that a random intercept was sufficient to capture the within-individual BMD measurement correlations. The estimated variances of the random intercept in both models can be found in Tables S1 and S2 in the supplementary materials.

The models in the following sections used a time-invariant individual variance for the predictor hormone. We also attempted to fit a linear time-varying individual variance component (see Simulation \ref{sec:sim2}) on the SWAN dataset, but encountered severe model convergence issues. We suspected that the signal in the SWAN dataset was not strong enough to support an individual time-varying variance specification. More discussion of this can be found in Section \ref{sec::discussion}.

\subsection{E2 Predictor Model}
In this model, we included interaction between the (lagged) estimated E2 residual and time to FMP when the BMD measurement was collected, and the interaction between E2 variability and time to FMP. Table \ref{tab:bmd_e2_coefs} displays the estimated posterior means and 95\% credible intervals for the outcome coefficients.

\begin{figure}
    \centering
        \includegraphics[scale=0.15]{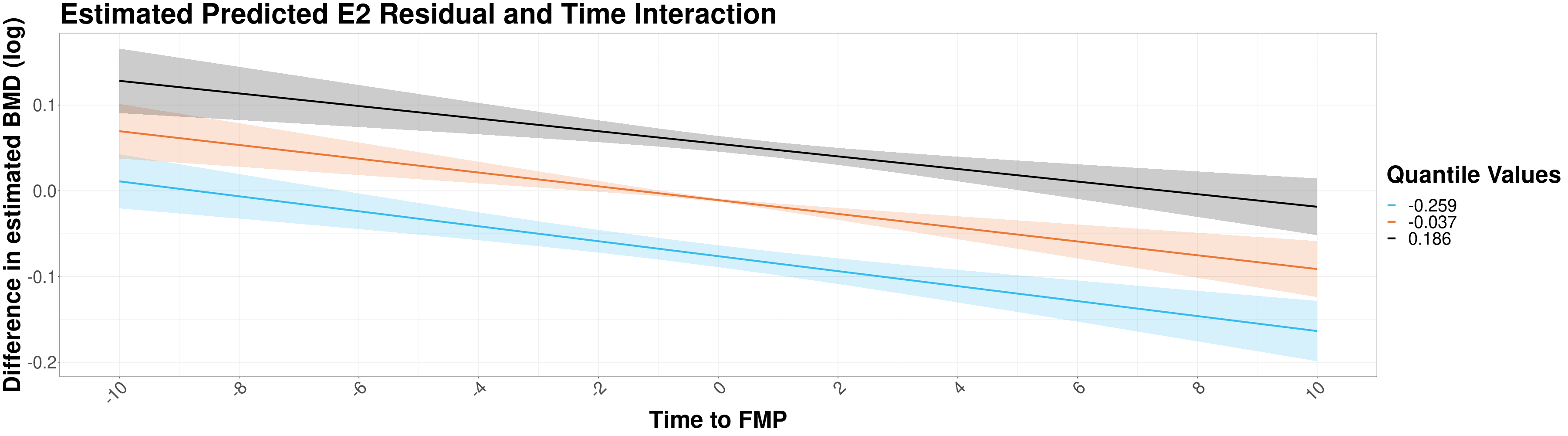}
 \caption{Plots of predicted E2 residuals and time interaction for the BMD outcome model. The solid lines represent the 25th, 50th, and 75th quantile values of the E2 variable, along with the prediction band for this value.}
 \label{fig::bmd_e2_interaction}
\end{figure}

The predicted E2 residual at the visit before FMP was significantly associated with BMD at FMP (i.e., when $t_{ij}$ = 0). The interpretation of the coefficient is that for women with a 1 unit (mg/l) higher predicted E2 residual at the visit before FMP, there was an average corresponding $(2^{0.3474}-1)\times 100\% = 22.8\% (19.1\%, 27.5\%)$ higher BMD at FMP. This effect was slightly moderated by time, since the coefficient for the interaction of predicted E2 and time is negative. However, the credible interval for the estimated interaction effect of time and predicted E2 contains 0, meaning that the estimated effect of E2 on BMD does not significantly change over time.  This is also evident from Figure \ref{fig::bmd_e2_interaction}, where the estimated slopes of the BMD trajectories do not change over the menopausal transition.

When predicted E2 is at the population average, an additional one year was associated with a $(1-2^{-0.0079})\times 100\% = -0.54\% (-0.77\%, -0.34\%)$ change in BMD. If we hold predicted E2 constant, then each additional year is associated with a $(1-2^{-0.0079}\times 2^{0.0031}) \times 100\% = -0.33\% (-0.70\%, 0.06\%)$ decrease in predicted BMD.

Higher E2 variability at FMP was negatively associated with BMD; a one unit increase in E2 variance was associated with a $(1-2^{-0.0238})\times 100\% = -1.6\% (-4.74\%, 1.43\%)$ change in BMD. However, since the 95\% CrIs contained 0, we cannot say that this relationship was statistically significant. The interaction term for E2 variance and time to FMP was also not significant. 

Finally, baseline BMI was positively associated with BMD, indicating that women with higher BMI tended, on average, to have $8.8\%$ higher BMD ($8.0\%, 9.74\%$), holding all else constant. Baseline age, however, was not significantly associated with BMD.

\subsection{FSH Predictor Model}

\begin{figure}%
    \centering
    \includegraphics[scale=0.14]{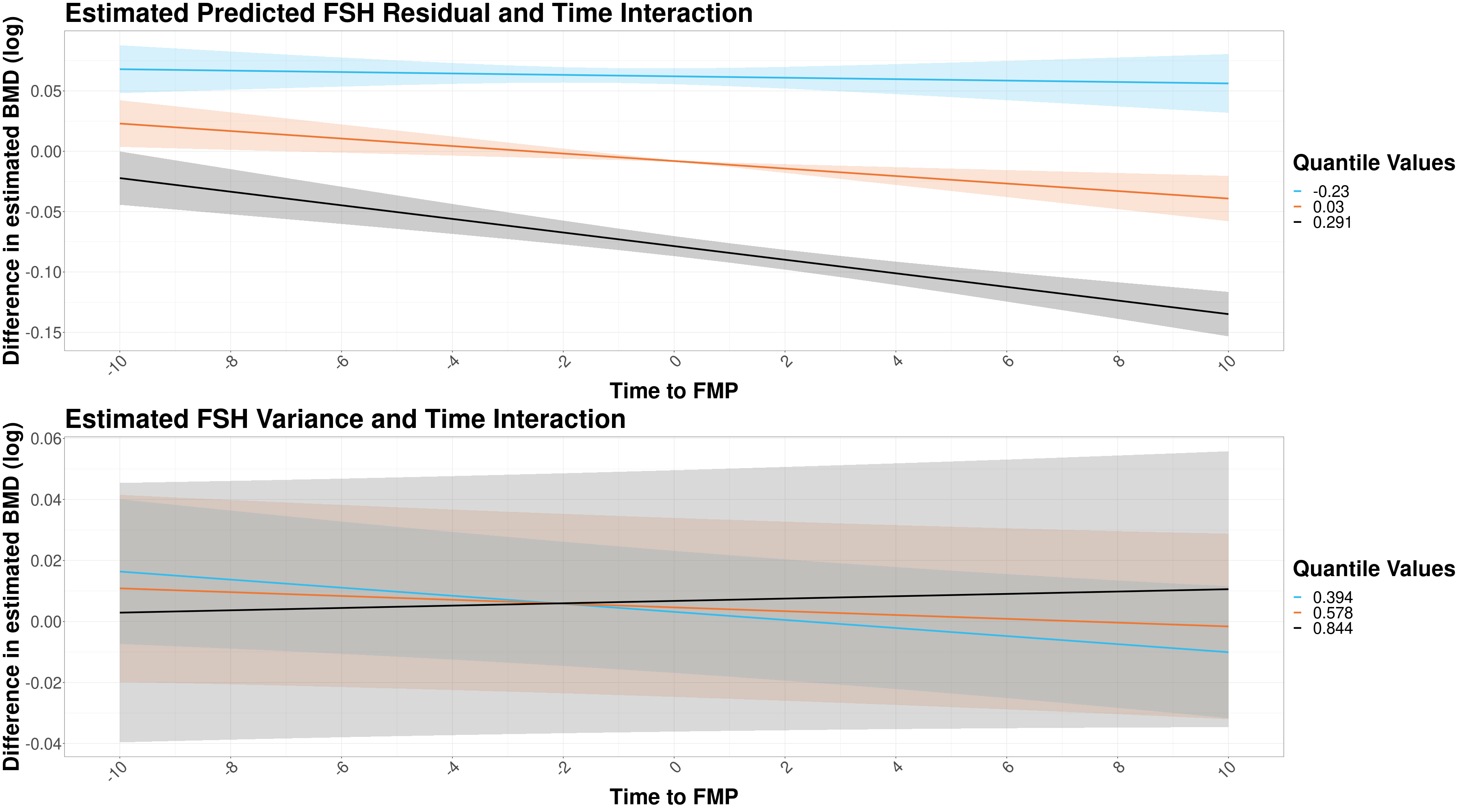}
        \caption{Plots of predicted FSH and time interaction (top figure) and FSH variance and time interaction (bottom figure) for the BMD outcome model. The solid lines represent the 25th, 50th, and 75th quantile values of the FSH variable, along with the prediction band for this value. We can see the moderating effect of the variance-time interaction term as the BMD residual trajectories converge at around 7 years post-FMP.}
            \label{fig::bmd_fsh_interaction}
\end{figure}

 Table \ref{tab:bmd_fsh_coefs} displays the estimated posterior means and 95\% credible intervals for the FSH model outcome coefficients. Predicted FSH residual at the visit before FMP was significantly associated with BMD at FMP (when $t_{ij} = 0$). For women with a 1 unit (pg/mL) higher predicted FSH residual at the visit before FMP, there was an average corresponding $(1-2^{-0.2700}) \times 100\% = -17.1\% (-18.7\%, -15.4\%)$ lower BMD at FMP. When predicted FSH is at the population average, an additional one year was associated with a $(1-2^{-0.0028}) \times 100\%  = -0.19\% (-0.33\%, -0.06\%)$ change in BMD. The FSH and time interaction was also significant and indicated that the association between mean FSH and BMD becomes amplified over time. If we hold predicted FSH constant, then each additional year was associated with a $(1-(2^{-0.0028} \times 2^{-0.0097}) \times 100\% = -0.86 \% (-1.15\%, -0.57\%)$ change in predicted BMD residual. This can also be seen in Figure \ref{fig::bmd_fsh_interaction}, where the estimated BMD trajectories tend to diverge after FMP.

FSH variability at FMP was not significantly associated with BMD. The interaction term, however, between variability and time to FMP was significant. Holding FSH variability constant, a one year increase in time to FMP is associated with a $(1-2^{-0.0028} \times 2^{0.0038})  = 0.07\% ( -0.25\%, 0.38\%)$ change in BMD. Since this interaction term has a positive sign, this indicates that the association of FSH variability with BMD is moderated over time. The bottom plot in Figure \ref{fig::bmd_fsh_interaction} shows this moderating effect where the estimated difference in BMD trajectories converge around 8 years post FMP. 

Finally, as in the E2 predictor model, baseline BMI was significantly associated with BMD. Women with higher BMI values tended to have higher BMD values, holding all else constant. This is consistent with the literature on associations between body size and BMD in adult women \citep{salamat_relationship_2013}. Baseline age was not significantly associated with BMD. 

\section{Conclusion}
\label{conclusion}
Our analyses found that higher mean E2 was associated with higher average predicted BMD, which supports the hypothesis that higher levels of E2 are protective against BMD loss, and thus that E2 supports bone health during the midlife. Conversely, higher FSH was associated with lower average BMD, which suggests that higher levels of FSH may negatively affect bone health in the midlife for women. These findings are consistent with the literature on the relationship between E2 and FSH levels and BMD outcomes in women during the midlife \citep{zaidi_fsh_2018,park_bone_2021, li_association_2023}. 

We also found that higher individual FSH variability over the MT (although not at FMP) was associated with slower declines in BMD. As shown in Figure \ref{fig::bmd_fsh_interaction}, higher FSH variability is associated with lower levels of BMD before FMP, but this effect is moderated over time during the MT. This suggests that women with low FSH variability may be at higher risk of bone loss post-FMP than they are before the MT. We did not find a similar relationship between BMD and E2 variability. This suggests that FSH variability (interacted with time) is more strongly predictive of BMD trajectories in menopausal women than E2 variability. To the best of our knowledge, this is the first time that \textit{individual-level} E2 and FSH variances have been used to predict repeatedly-measured BMD outcomes in women. These findings should motivate further research into the role of FSH variability on BMD declines to more fully understand the relationship between individual variances of hormones and women’s BMD trajectories. 

\section{Discussion}
\label{sec::discussion}

Maintaining bone health in women is particularly important during the midlife, since bone resorption begins to outpace bone formation as women approach menopause \citep{demontiero_aging_2012}. Given that declines in BMD accelerate during the menopausal transition, it is particularly critical to promote bone health during this lifestage \citep{finkelstein_bone_2008}. Additionally, the menopausal transition is a period of time with large mean changes in sex hormone levels and increased variability of these hormones. Although much research has evaluated associations between sex hormone levels and BMD, there is still a lack of understanding of how hormone variabilities may affect BMD loss in women undergoing menopause. Little is known in the current literature about the roles of individual-level E2 and FSH variances in predicting bone health, although previous analyses have found associations between individual E2 and FSH variances and other health outcomes (such as risk of hot flash) \citep{jiang_joint_2015, chen_joint_2023}. Our study contributes to this important, but not yet extensive, literature that connects individual-level hormone variances and health outcomes across the MT.

In addition to women's bone health during the midlife, we believe that modeling variabilities as predictors can be of wide interest in many applications. Intra-individual heart rate variability and cognitive variability are two measures that are known to be highly predictive of adverse health outcomes \citep{fang_heart_2020, santos-de-araujo_inter-_2024, lin_intra-individual_nodate}. In addition to these public health and psychology applications, sociology research has indicated that individual variability of measures like subjective well-being and life satisfaction can be highly predictive \citep{gadermann_investigating_2007, quek_understanding_2025}. These measures of variability are commonly calculated first and then used to predict a health outcome such as dementia onset or mortality. Our joint modeling approach provides a way to carry the uncertainty of the variance estimates into the prediction of the outcome, which will reduce the bias and uncertainty in the final predictions, which can open up use of subject-level variability across a wide range of applications in health sciences. 

Our analysis had a few limitations. As noted in the introduction, BMD has a nonlinear rapid decrease as women approach the menopausal transition, and then stabilizes around 2-5 years after FMP. Originally, we attempted to fit an outcome model with a linear spline on time to FMP, with knots at 2 years before and after FMP. This model had trouble converging for both hormone predictors. In particular, the coefficients that represented the mean and variance interacted with the spline on time to FMP failed to converge. To address this, we decided to remove the nonlinear BMD trend by using the individual BMD residuals as the main outcome of interest; this left subject-level trends that could be modeled linearly. 
Another potential limitation is selection bias resulting from the reduction to an analytic sample of 972 from the full cohort of 3,302 women.  However, most of this reduction was due to the fact that two of the seven study sites were excluded because of the lack of DEXA technology to record BMD measurements, or because of women failing to maintain menstruation calendars in sufficient detail to pinpoint FMP date. While we are confident that the resulting analytic sample is a reasonably random subset of the original, we cannot exclude the possibility that selection bias impacts inference
Finally, as mentioned in Section \ref{sec::swan_application}, when we attempted to fit a linear time-varying variance on the SWAN dataset, the MCMC chains for the variance parameters in both models failed to converge to a single value. This also occurred when we fit only the longitudinal predictor model to the hormone data, which further suggests that there was not enough signal in the hormone data to support a time-varying variance trend. One reason for this could be due to the relatively small number of available observations per individual, as well as observations being collected annually. It may be easier to capture a time-varying variance trend in observations that are collected closer together over time. Since the individual mean trajectories are also estimated in our model, we hypothesize that there was not enough remaining signal to adequately estimate a time-varying variance for each individual. 

\paragraph{Future Work} 

Even though BMD is the standard metric of choice used to evaluate bone health, some have cautioned against solely relying on BMD to measure bone health. \cite{prentice_uncritical_1994} argued that BMD should not be used in epidemiological studies. Since the formula for BMD assumes a constant proportional relationship between bone area and BMC (bone mineral content), this can lead to spurious correlations between BMD and health outcomes when the relationship between BMC and bone area is not directly proportional. ``If BMD is used when the relationship between BMC and BA [bone area] is not one of simple direct proportion, part of its variation within a population will be due to differences in bone size between individuals." Now that we have looked at the association between hormone variability and BMD, the next question of interest would be to understand if these associations are also present with women’s BMC and bone area trajectories.

The SWAN study has collected women's femoral neck BMC (g) measurements and femoral neck area (cm$^2$) measurements visits, but due to changes in the DXA collection machines over the course of the study, these measurements are not yet appropriately calibrated for longitudinal analyses. When the calibrated measurements become available, we plan to apply our model to BMC trajectories and bone area trajectories. Another possible extension of the model would be to simultaneously model both BMC and bone area trajectories as a multivariate outcome, rather than separately analyzing each variable. 


An interesting future area of research could be to model complex formulation of the individual variances. In particular, decomposing individual time-varying variances into short-term and long-term trends may be of scientific interest for epidemiologists and physicians. Exploring these higher-order trends and obtaining sufficiently stable estimates of the variance trends will require a dataset with either a stronger signal for time-varying trends or a larger number of observations per individual in the longitudinal predictors. 

Finally, our joint model is specified for one longitudinal predictor marker, but this could be extended to multiple longitudinal markers by specifying $\Bb_i = [\bb_{i1},...,\bb_{iQ}]^{\transp}$, where $Q$ is the number of markers, and a $Q \times Q$ variance-covariance matrix $\Sb_i$ for each individual. This extension would have several considerations. The computation cost of estimating $\Sb_i$ would grow non-linearly as $Q$ increases. Additionally, if we wanted to model both a mean regression and covariance regression (e.g. time-varying covariance matrices) in the predictor model, then this would further increase the computational burden of estimating the model.

\bmhead{Supplementary information}

We provide supplementary material containing additional simulation results, as well as model checks and diagnostics in the accompanying file. Additionally, the R code to run the model and replicate the simulation studies can be accessed at the following GitHub repository: https://github.com/realirena/jelo/. Access to the SWAN dataset is restricted due to its sensitive nature, but can be applied for at the following link: https://www.swanstudy.org/swan-research/data-access/. 

\bmhead{Acknowledgments}

The content of this article is solely the responsibility of the authors and does not necessarily represent the official views of the NIA, NINR, ORWH or the NIH. This research also was supported in part through computational resources and services provided by Advanced Research Computing (ARC), a division of Information and Technology
Services (ITS) at the University of Michigan, Ann Arbor. A preprint of this paper  \citep{chen_joint_2023} has been uploaded to the Arxiv open-access archive and can be accessed at the following link: 
https://doi.org/10.48550/arXiv.2309.08000. The contents of this paper also appear in Chapter 3 of the author's PhD dissertation \citep{chen_thesis_2023}. 

\underline{Clinical Centers:}  \textit{University of Michigan, Ann Arbor – Carrie Karvonen-Gutierrez, PI 2021 – present, Siobán Harlow, PI 2011 – 2021, MaryFran Sowers, PI 1994-2011; Massachusetts General Hospital, Boston, MA – Sherri‐Ann Burnett‐Bowie, PI 2020 – Present; Joel Finkelstein, PI 1999 – 2020; Robert Neer, PI 1994 – 1999; Rush University, Rush University Medical Center, Chicago, IL – Imke Janssen, PI 2020 – Present; Howard Kravitz, PI 2009 – 2020; Lynda Powell, PI 1994 – 2009; University of California, Davis/Kaiser – Elaine Waetjen and Monique Hedderson, PIs 2020 – Present; Ellen Gold, PI 1994 - 2020; University of California, Los Angeles – Arun Karlamangla, PI 2020 – Present; Gail Greendale, PI 1994 - 2020; Albert Einstein College of Medicine, Bronx, NY – Carol Derby, PI 2011 – present, Rachel Wildman, PI 2010 – 2011; Nanette Santoro, PI 2004 – 2010; University of Medicine and Dentistry – New Jersey Medical School, Newark – Gerson Weiss, PI 1994 – 2004; and the University of Pittsburgh, Pittsburgh, PA – Rebecca Thurston, PI 2020 – Present; Karen Matthews, PI 1994 - 2020.} 

\underline{NIH Program Office:}  \textit{National Institute on Aging, Bethesda, MD – Rosaly Correa-de-Araujo 2020 - present; Chhanda Dutta 2016- present; Winifred Rossi 2012–2016; Sherry Sherman 1994 – 2012; Marcia Ory 1994 – 2001; National Institute of Nursing Research, Bethesda, MD – Program Officers.}

\underline{Central Laboratory:}   \textit{University of Michigan, Ann Arbor – Daniel McConnell} (Central Ligand Assay Satellite Services). 

\underline{Coordinating Center:}   \textit{ University of Pittsburgh, Pittsburgh, PA – Maria Mori Brooks, PI 2012 - present; Kim Sutton-Tyrrell, PI 2001 – 2012; New England Research Institutes, Watertown, MA - Sonja McKinlay, PI 1995 – 2001.}

Steering Committee:	Susan Johnson, Current Chair,  Chris Gallagher, Former Chair 
\textbf{We thank the study staff at each site and all the women who participated in SWAN.} 

\section*{Declarations}

The Study of Women's Health Across the Nation (SWAN) has grant support from the National Institutes of Health (NIH), DHHS, through the National Institute on Aging (NIA), the National Institute of Nursing Research (NINR) and the NIH Office of Research on Women’s Health (ORWH) (Grants U01NR004061; U01AG012505, U01AG012535, U01AG012531, U01AG012539, U01AG012546, U01AG012553, U01AG012554, U01AG012495, and U19AG063720). This work also was supported by National Institute on Aging Grant 1-R56-AG066693.  

The authors declare no potential conflict of interests.


\maketitle

\bibliography{jelo_bib}


 \begin{table}

\begin{tabular*}{\textwidth}{@{\extracolsep{\fill}}llrrr@{\extracolsep{\fill}}}
\toprule
\textbf{{Variable}}&\textbf{Statistic}&\textbf{Value}&\textbf{{n}}\\
\midrule
\midrule
\textit{Longitudinal Predictor} & Mean/SD &&\\
E2 Residuals & &  -0.01 (1.15) &8,383 \\
FSH Residuals & &  0.02 (0.90) &8,383 \\
\addlinespace 
\textit{Health Outcome} & Mean/SD &&\\
BMD Residuals & & 0.00 (0.23) & 8,383 \\
\addlinespace 
\textit{Adjusted Covariates} & Mean/SD &&\\
Baseline BMI & & 27.28 (6.83) & 8,383\\
Baseline Age & & 46.29 (2.60) &8,383 \\
\bottomrule
\end{tabular*}
\caption{Descriptive statistics of the dataset, based on 974 individuals in the Study of Women’s Health Across the Nation.}
 \label{tab:swan_descriptive_stats}
\end{table} 

\begin{table}
\centering
\begin{tabular*}{\textwidth}{@{\extracolsep{\fill}}llrrr@{\extracolsep{\fill}}}
\toprule
{Truth}&{Model}&{Bias}&{Coverage (\%)}&{Average Interval Length}\\
\midrule
$\beta_{0}$ = 2 & \textbf{JELO} &\textbf{0.00} & \textbf{95.5} & \textbf{0.28}\\
                &TSLMM &-1.63 & 53.5 & 0.44\\
                &TSLO &  0.05 & 0.0 & 0.43\\
\addlinespace
$\beta_{1}$ = -0.1 & \textbf{JELO} &\textbf{0.00} & \textbf{95.5} & \textbf{0.06}\\
                &TSLMM &-0.09 & 2.0 & 0.09\\
                 &TSLO &0.00 & 44.0 & 0.07\\
\addlinespace
$\beta_{2}$ = -1 & \textbf{JELO} & \textbf{0.00} & \textbf{95.0} & \textbf{0.22}\\
                &TSLMM &0.43 &  57.0 & 0.20\\
                &TSLO &-0.02 & 43.5 & 0.21 \\
 \addlinespace
$\beta_{3}$ = -0.75 & \textbf{JELO} &\textbf{0.00} & \textbf{94.5} & \textbf{0.26}\\
                &TSLMM & 0.58  & 0.0& 0.06\\
                 &TSLO &0.00 & 72.5 & 0.38 \\
\addlinespace
$\beta_{4}$ = -0.5 & \textbf{JELO}& \textbf{0.00} & \textbf{96.5} & \textbf{0.04}\\
             &TSLMM & -0.04 & 0.0& 0.08 \\
             &TSLO & -0.003 & 43.5  & 0.03 \\
\addlinespace
$\beta_{5}$ =  0.2 & \textbf{JELO} & \textbf{0.00} & \textbf{94.0} & \textbf{0.26}\\
                &TSLMM &0.01 & 0.0 & 0.07\\
                &TSLO &0.00 & 52.5 & 0.15\\
\bottomrule
\end{tabular*}
\caption{ Simulation I: bias, coverage, and mean 95\% credible interval (or confidence interval) length across 200 simulation replicates.}
\label{tab::sim1_outcome_betas}
\end{table}

\begin{table}[t]
\tabcolsep=0pt
\centering
\begin{tabular*}{\textwidth}{@{\extracolsep{\fill}}lllrrr@{\extracolsep{\fill}}}
\toprule
{Truth}&{Scenario}&{Model}&{Bias}&{Coverage (\%)}&{Average Interval Length}\\
\midrule
$\beta_{0}$ = 2 & HV & JELO &0.00 & 96.5 & 0.12\\
        & HV  &   JELO (CV) &0.05 & 68.5 & 0.12\\
        & LV & JELO &0.00 & 95.0 & 0.24\\
        & LV  & JELO (CV) & 0.00 & 93.0 & 0.24 \\
\addlinespace
$\beta_{1}$ = -1.5  & HV  & JELO &0.00 & 91.0 & 0.08\\
& HV   & JELO (CV) &0.00 & 93.5 & 0.08\\
 & LV & JELO &0.00& 92.5 & 0.10\\
 & LV  &   JELO (CV) &0.00  &91.0 & 0.10 \\
\addlinespace
$\beta_{2}$ = 0.25 & HV  & JELO &0.00 & 94.5 & 0.15\\
& HV  &JELO (CV)  &-0.20 & 0.5 & 0.11\\
 & LV & JELO &0.00 & 95.0 & 0.19\\
 & LV  &   JELO (CV) &0.00 & 95.4 & 0.19\\
 \addlinespace
$\beta_{3}$ =  1 & HV  & JELO &0.00 & 98.5 & 0.18\\
& HV &JELO (CV)  &-0.05 & 82.5 & 0.18\\
 & LV & JELO &-0.01 & 97.5 & 0.06\\
 & LV  &  JELO (CV) & 0.00 & 97.0 & 0.06\\
\addlinespace
$\beta_{4}$ =0.75 & HV &  JELO &0.00 & 95.5 & 0.05\\
& HV &JELO (CV)  &0.01 & 82.5 & 0.05\\
& LV & JELO &0.00 & 95.0 & 0.27\\
 & LV  &   JELO (CV) & -0.01 & 96.4 & 0.26\\
\addlinespace
$\beta_{5}$ = -0.10 & HV & JELO &0.00 & 96.0 & 0.12\\
& HV &JELO (CV) &0.29 & 0.0 & 0.10\\
 & LV & JELO &0.00 & 99.0 & 0.06\\
 & LV  &  JELO (CV) &0.01 & 96.4 & 0.13\\
\bottomrule
\end{tabular*}
\caption{Simulation II: bias, coverage, and 95\% credible interval length across 200 simulation replicates for each data scenario (CV = constant variance) for selected parameters. }
\label{tab::sim2_outcome_betas} 
\end{table}

\begin{table}

\centering
\renewcommand{\arraystretch}{0.2}
\begin{tabular}{r r r r r}
\toprule
\multicolumn{1}{c}{Variable}&\multicolumn{1}{c}{Post. Mean}&\multicolumn{1}{c}{95\% CrI}\tabularnewline
\midrule
\tabularnewline
\textbf{Predicted E2} & \textbf{29.46} & (\textbf{25.25},\textbf{35.00)} \\
E2 Var. & -2.38 & (-6.76, 2.05)\\
\tabularnewline
\textbf{Time to FMP} & \textbf{-0.79} & (\textbf{-1.12}, \textbf{-0.49}) \\
\tabularnewline
\textbf{Time to FMP x Predicted E2} & \textbf{0.31} & \textbf{(0.12, 0.58)} \\
\tabularnewline
Time to FMP x E2 Var. & 0.21 & (-0.01,0.46 )\\
\textbf{BMI} & \textbf{12.21} & (\textbf{11.00, 13.41})\\
\tabularnewline
Age & 0.11 & (-0.36, 0.57)\\
\tabularnewline
\bottomrule
\end{tabular}
\caption{Estimated posterior means and 95\% credible intervals for the E2-BMD model with time interactions. All estimated posterior means and 95\% CrI values have been multiplied by $10^{2}$.}
\label{tab:bmd_e2_coefs} 
\end{table}

\begin{table}
\centering
\renewcommand{\arraystretch}{0.2}
\begin{tabular}{r r r r r}
\toprule
\multicolumn{1}{c}{Variable}&\multicolumn{1}{c}{Post. Mean}&\multicolumn{1}{c}{95\% CrI}&\tabularnewline
\midrule
\tabularnewline
\textbf{Predicted FSH} & \textbf{-27.00} & (\textbf{-29.83, -24.25}) \\
FSH Var. &  0.80 & (-4.29, 5.95) \\
\tabularnewline
\textbf{Time to FMP} & \textbf{-0.28} & (\textbf{-0.48, -0.09}) \\
\tabularnewline
\textbf{Time to FMP x Predicted FSH} & \textbf{-0.97} & (\textbf{-1.20, -0.74})\\
\tabularnewline
\textbf{Time to FMP x FSH Var.} & \textbf{0.38} & (\textbf{0.12, 0.64})  \\
\textbf{BMI} & \textbf{9.86} & (\textbf{8.58, 11.24})\\
\tabularnewline
Age & -0.22 & (-0.70, 0.29)\\
\tabularnewline
\bottomrule
\end{tabular}
\caption{Estimated posterior means and 95\% credible intervals for the FSH-BMD model with time interactions. All estimated posterior means and 95\% CrI values have been multiplied by $10^{2}$.}
 \label{tab:bmd_fsh_coefs}
\end{table} 

\end{document}